\documentclass{aa1}  
\usepackage{graphicx}
\usepackage{threeparttable}
\usepackage{txfonts}
\usepackage{xcolor, color}
\definecolor{xlinkcolor}{cmyk}{1,1,0,0}
\usepackage{natbib}
\bibpunct{(}{)}{;}{a}{}{,} 
\usepackage[bookmarks=true, pdfnewwindow=true, colorlinks=true, linkcolor=xlinkcolor, citecolor=xlinkcolor, filecolor=xlinkcolor, urlcolor=xlinkcolor, final=true]{hyperref}

\begin{document} 

   \title{An isotropic full-sky sample of optically selected blazars}
   \author{Maria Kudenko
          \inst{1,2}\fnmsep\thanks{Corresponding author: \email{kudenko.ma19@physics.msu.ru}}
          \and
          Sergey Troitsky\inst{1,2}
          } 
   \institute{Institute for Nuclear Research of the Russian Academy of Sciences, 60th October Anniversary prospect~7a, 117312 Moscow, Russia                  \and
             Physics Department, Lomonosov Moscow State University, 1-2 Leninskie Gory,  Moscow 119991, Russia}
   \date{Submitted on December 12, 2023}
 
  \abstract
   {Various high-energy phenomena in the Universe are associated with blazars, which are powerful active galaxies with jets pointing at the observer. Novel results relating blazars to high-energy neutrinos, cosmic rays, and even possible manifestations of new particle physics, often emerge from statistical analyses of blazar samples, and uniform sky coverage is important for many of these studies. }
   {Here, we construct a uniform full-sky catalog of blazars selected by their optical emission.}
   {We defined the criteria of isotropy, making a special effort to cover the Galactic plane region, and compiled an isotropic sample of blazars with GAIA optical magnitudes of $G<18^{\rm m}$, corrected for Galactic absorption. The sources were taken from full-sky samples selected by parsec-scale radio emission or by high-energy gamma-ray flux, both of which are known to efficiently select blazar-like objects.}
   {We present a catalog of 651 optically bright blazars, uniformly distributed in the sky, together with their radio, optical, X-ray, and gamma-ray fluxes, and an isotropic sample of 336 confirmed BL Lac type objects.}
   {This catalog may be used in future statistical studies of energetic neutrinos, cosmic rays, and gamma rays.}

   \keywords{catalogs --
                galaxies: active --
                BL Lacertae objects: general --
                quasars: general --
                astroparticle physics
               }

   \maketitle
%

\section{Introduction}
\label{sec:intro}
Multimessenger astronomy is currently exploring its extreme high-energy limits. Astrophysical neutrinos with energies from TeVs to PeVs have been detected by the IceCube \citep{IceCubeFirst26,IceCube-HESE-2020}, ANTARES \citep{ANTARES2019ICRC}, and Baikal-GVD \citep{Baikal-diffuse} experiments. New data are coming from the KM3NeT \citep{KM3NeT-experiment} observatory, while future detectors, P-ONE \citep{P-ONE}, IceCube-GEN2 \citep{IceCube-Gen2}, NEON \citep{NEON-1km3-dense}, TRIDENT \citep{TRIDENT-8km3}, and HUNT \citep{HUNT-30km3}, are being developed. Neutrinos with higher energies $\gtrsim 10^{17}$~eV will be detected and studied in the ARA \citep{ARA}, ARIANNA \citep{ARIANNA}, RNO-G \citep{RNO-G}, GRAND \citep{GRAND}, JEM-EUSO \citep{JEM-EUSO}, and POEMMA \citep{POEMMA} experiments, and others. Cosmic rays with even higher energies are routinely being detected by the Pierre Auger Observatory \citep{Auger-experiment} and the Telescope Array \citep{TelescopeArray:2008toq} experiments, and are observed up to primary energies of several times $10^{20}$~eV \citep[e.g.][]{TA-Science}.

Despite the progress on the experimental side, the origin of these extremely energetic particles remains elusive (see e.g.,\ 
\citet{Meszaros-rev,ST-UFN,ST-UFN2} for neutrinos and \citet{Anchordoqui-rev,Kachelriess-Semikoz-rev,Kachelriess-rev2022,Kuznetsov:2023jfw} for cosmic rays). Data are scarce at these high energies; in addition, low-directional-accuracy and non-astrophysical backgrounds in neutrino detectors, as well as magnetic deflections of charged cosmic rays, make it very hard to determine sources of the energetic particles on an event-by-event basis. To find populations of potential astrophysical sources, one should rely on statistical analyses, which attempt to relate catalogs of sources to samples of observed events under certain assumptions. 

A combination of small and large scale anisotropy searches is among these statistical approaches, so that a catalog of putative sources is correlated with event directions at small angular scales, while the overall distribution of events in the sky provides for additional constraints on their origin. For studies of this kind, having an isotropic catalog of sources is important, as we illustrate here with two examples.

Firstly, making use of an isotropic sample is important when one needs to separate between local and distant effects contributing to observations. An example is provided by the puzzling directional correlations of BL Lac type objects (BL Lacs) with ultra-high-energy cosmic rays detected by the High Resolution Fly's Eye experiment \citep[HiRes;][]{Gorbunov:2004,HiRes:HiRes} and found at angular scales comparable to the HiRes angular resolution. In one of the scenarios explaining these anomalous associations \citep{FRT}, it was predicted that the overall distribution of the correlated events should follow the local large-scale structure of the Universe, and this effect was indeed found \citep{ST-anisotropy}. However, since the sample of BL Lacs by \citet{Veron}, used there, was not isotropic, the result might be subject to unspecified biases.

Secondly, the isotropy of the sample may be crucial when there are two populations of sources contributing to the total flux of energetic particles. An example, discussed by \citet{neutgalaxy}, includes the sources of high-energy neutrinos, which include both distant active galactic nuclei and our Milky Way. The null hypothesis assumed for the search of neutrino correlations 
with any population of sources is the isotropic distribution of neutrino arrival directions, while the presence of the Galactic component violates this assumption. 

Blazars, defined as active galactic nuclei with relativistic jets pointing to the observer, are among the most probable sources of energetic particles in the Universe. This is because they provide the physical conditions for the acceleration of charged hadrons to extreme energies, while the Doppler enhancement associated with the jet geometry results in higher observed fluxes. There exist observational indications that other astrophysical objects are associated with neutrino events, including Seyfert galaxies \citep{NGC1068}, tidal disruption events \citep{TDE}, and the Milky Way \citep{neutgalaxy,ANTARES-ridge,IceCube-gal-Science}, so the neutrino sky is richer than one thought several years ago \citep[for a review, see][]{ST-UFN2}. However, numerous studies \citep[e.g.,][]{neutradio1,neutradio2,Elisa:blazars,Anna:blazars,Buson1,Buson2} indicate that blazars provide a large contribution to the full-sky astrophysical neutrino fluxes. Notably, while the contribution of gamma-ray loud blazars is observationally constrained from above \citep[e.g.,][]{GammaBlazarLimits}, the significance of associations with radio-selected ones grows \citep{neutradio2023} with the number of neutrino events observed after the first claims. The contribution of blazars to the cosmic ray flux is worse understood now, but both theoretical arguments and observed associations discussed above select BL Lacs, a subclass of blazars defined below in Sec.~\ref{sec:construction:bll}, as a possible source population. 

Here, we construct, making use of various published data, an isotropic full-sky sample of blazars with Gaia $G$-band magnitudes, corrected by the Galactic absorption, $G_{\rm corr}<18^{\rm m}$. In Sec.~\ref{sec:criteria}, we recall the main sources of anisotropy in previously used data samples and formulate our technical criteria, defining which sample is sufficiently isotropic for our purposes. Section~\ref{sec:construction} describes the procedures we followed to construct the catalog, as well as the data we used. Section~\ref{sec:catalog} contains a description of the catalog and presents two isotropic samples, one of all blazars and another one of confirmed BL Lacs. Full catalogs are available online in a machine-readable format. We briefly conclude in Sec.~\ref{sec:concl}.

\section{Criteria of isotropy}
\label{sec:criteria}
\begin{figure}
   \centering
   \includegraphics[width=\hsize]{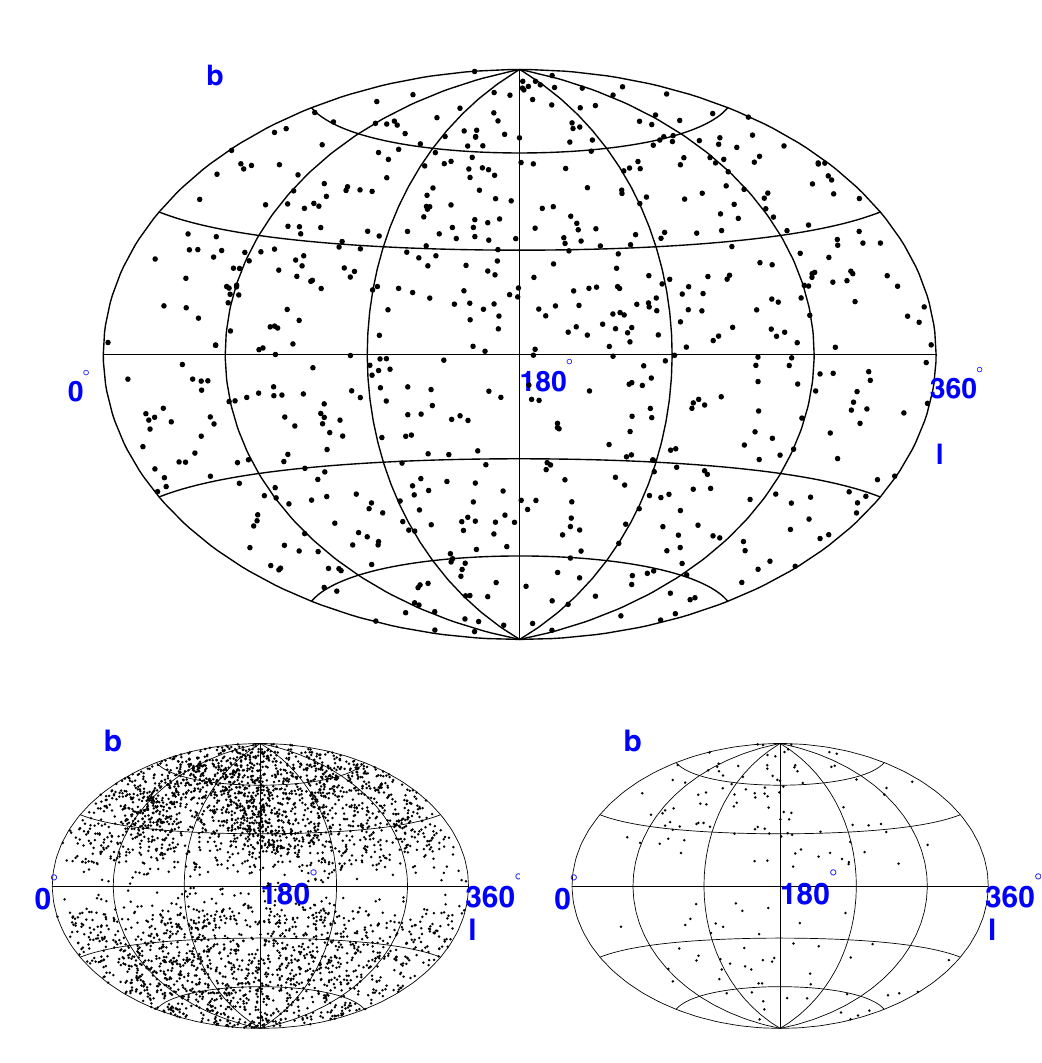}
      \caption{Sky maps (Galactic coordinates) illustrating (an)isotropy of various blazar samples. Top: The catalog presented in this study. Bottom left: The 5th edition of the Roma-BZCAT catalog \citep{BZcat} used e.g.,\ by \citet{Buson1,Buson2}. Bottom right: The sample of optically selected blazars from \citet{Veron} used by \citet{Gorbunov:2004}.}
         \label{Fig:Maps}
   \end{figure}

Two main sources of large-scale anisotropy of available catalogs are the uneven coverage of the sky by various instruments and the Galactic absorption. It should be noted that the latter not only biases the sensitivity to detect particular sources but also limits the optical spectroscopic studies required for identification of blazars. The patchy structure of multi-instrument compilations and the lack of objects around the Galactic plane are seen, for two popular blazar catalogs, in the bottom panel of Fig.~\ref{Fig:Maps}.
The effect of the identification problem on the Galactic anisotropy is illustrated in Fig.~\ref{Fig:galactic_no_cut}: 
\begin{figure}
   \centering
   \includegraphics[width=\hsize]{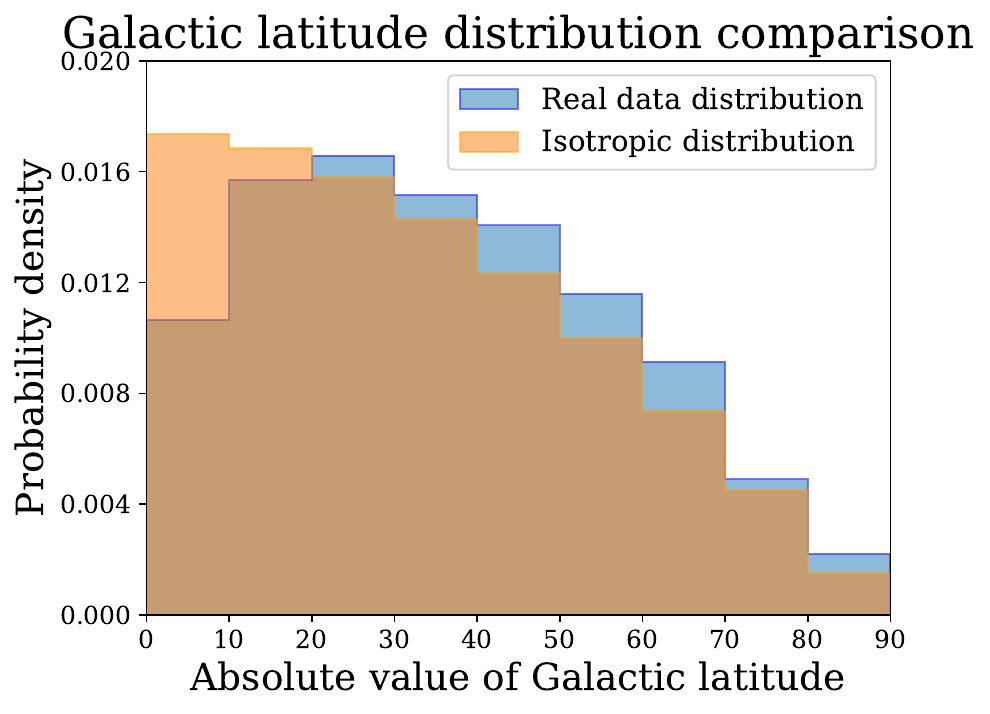}
      \caption{Normalized Galactic latitude distribution of blazar-like objects from the Fermi Point Source Catalog \citep[4FGL,][]{4FGL} in comparison with the isotropic distribution.
              }
         \label{Fig:galactic_no_cut}
   \end{figure}
while the Milky Way is transparent for high-energy gamma rays, the distribution of identified Fermi blazars in the Galactic latitude is far from isotropic.

Our approach to constructing the isotropic sample, which is presented in the top panel of Fig.~\ref{Fig:Maps} for comparison, was to start with flux-limited samples covering various classes of blazars and to change the corresponding limit toward higher fluxes until the sample satisfied our predefined criteria of isotropy: brighter objects are observed more uniformly across the sky. We started with the formulation of these criteria.
   
\subsection{Galactic-latitude distribution}
\label{sec:criteria:latitude}
To control the Galactic-plane dip in the distribution of blazars, we used the distribution of absolute values of Galactic latitude $b$ of sources and compared it with the distribution of an isotropic set. To this end, we used the Kolmogorov--Smirnov test. 
A sample is considered isotropic over the Galactic latitude if the Kolmogorov--Smirnov probability that the data follow the isotropic distribution exceeds 10\%. This number was selected a priori in order to keep the balance between the isotropy and the depletion of the sample.

\subsection{Medium-scale isotropy}
\label{sec:criteria:sphere}
To test whether a sample of blazars was isotropic on medium angular scales, we constructed the following function defined for each point on the celestial sphere. We fixed a certain angle, $\theta$, and counted the number of sources within the cone with the opening angle $\theta$ around a given point; this integer number was treated as the function value. These  values were calculated for the sets of points represented by HEALPix pixel centers as in \cite{HEALPix} with the resolution parameter, $N_{\rm side}$. 

For the isotropic distribution of sources, the number of objects should follow the binomial distribution with the average 
\begin{equation} 
\bar{N}=N_0\cdot\left(\sin \frac{\theta}{2}\right)^2,      
\end{equation} 
where $N_0$ is the total amount of blazars in the sample, and $\theta$ is the opening cone angle. Again, we compared the real distribution of the number of objects with the binomial distribution by means of the Kolmogorov--Smirnov test and accepted the sample as isotropic if the Kolmogorov--Smirnov probability exceeded 10\%. This test was performed for three different sets of directions with $N_{side} = 2, 4, 16$ and cone angles $\theta$ = $3^{\circ}$, $13^{\circ}$, $23^{\circ}$, respectively. 

\section{Construction of the catalog}
\label{sec:construction}

\subsection{Starting samples}
\label{sec:construction:start}
Blazars are defined by the presence of a relativistic jet pointing to the observer within several degrees from the line of sight. The most general criterion of such a jet is the compact, parsec-scale radio emission, detected with the very long baseline interferometry (VLBI). We therefore used the catalog of VLBI selected sources as one of our starting samples. However, some high-frequency peaked blazars are comparably weak in the radio band, so we supplemented the VLBI sample with the sample of identified blazars detected in energetic gamma rays by the Fermi Large Area Telescope (LAT). Both samples, described below, are commonly accepted as efficiently selecting blazars. We note that both in radio and in gamma rays, the Galactic absorption is small, and anisotropies in the catalogs are entirely related to the observational biases discussed at the beginning of Sec.~\ref{sec:criteria}.

\subsubsection{The radio VLBI sample}
\label{sec:construction:start:RFC} 
\begin{figure}
   \centering
   \includegraphics[width=\hsize]{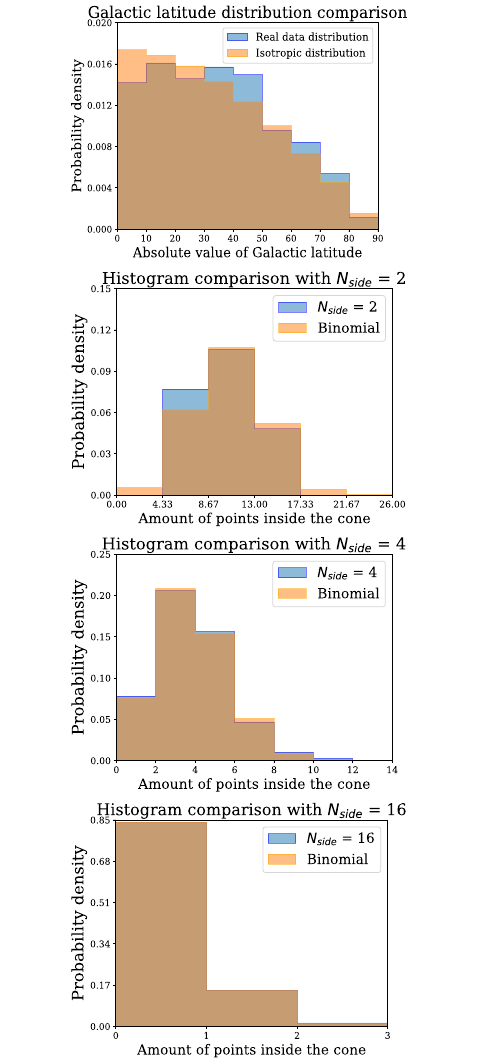}
      \caption{Normalized histograms for the VLBI-selected subsample with $G_{\rm corr}<18^{\rm m}$ and $F_{\rm 8~GHz}> 0.41$~Jy. Top: The distribution in the absolute value of the Galactic latitude in comparison with the isotropic one for the VLBI sample. Other plots represent the distributions described in Sec.~\ref{sec:criteria:sphere} for three values of $N_{\rm side}$, marked in the plots. 
              }
         \label{Fig:RFC_tests}
   \end{figure}

The Radio Fundamental Catalogue (\url{http://astrogeo.org/sol/rfc/rfc_2023b/}) provides positions, maps, and estimates of the correlated flux density for over 20,000 extragalactic radio sources produced by analysis of all available VLBI observations \citep{RFC1,RFC2,RFC3,RFC4,RFC5,RFC6}. We removed sources identified with non-blazar-type objects from the catalog by making use of the associations established in the OCARS \citep[][\url{http://www.gaoran.ru/english/as/ac_vlbi/}]{OCARS} or Simbad \citep{Simbad} databases.  Before applying our criteria of isotropy, we selected the objects with $G_{\rm corr}<18^{\rm m}$, as is described below in Sec.~\ref{sec:construction:GAIA}. To satisfy our criteria of isotropy, we tuned the minimal 8-GHz flux density $F_{\rm 8~GHz}$ in the sample. The criteria described in Sec.~\ref{sec:criteria:latitude}, \ref{sec:criteria:sphere} are satisfied for $F_{\rm 8~GHz}> 0.41$~Jy, leaving only 261 objects in the set. The resulting distributions for the set are presented in Fig.~\ref{Fig:RFC_tests}. 

\subsubsection{The gamma-ray sample selected}
\label{sec:construction:start:4FGL}
\begin{figure}
   \centering
   \includegraphics[width=\hsize]{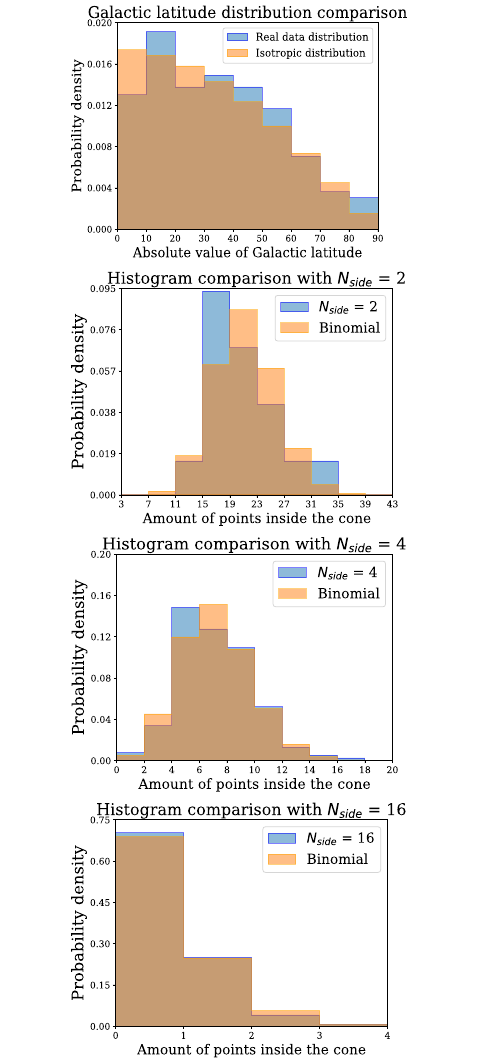}
      \caption{Same as Fig.~\ref{Fig:RFC_tests}, but for the 4FGL sample with $G_{\rm corr}<18^{\rm m}$ and $F_{\rm 1-100~GeV}> 3.8\cdot10^{-10}$~cm$^{-2}$s$^{-1}$.
              }
         \label{Fig:Fermi_tests}
   \end{figure}

The Fermi Point Source Catalog (4FGL) is a catalog of high-energy gamma-ray sources detected by the LAT on the Fermi Gamma-ray Space Telescope \citep{4FGL}. We used the identifications provided in the catalog and selected compact steep-spectrum quasars (css), BL Lacs (bll), flat-spectrum radio quasars (fsrq), and soft-spectrum radio quasars (ssrq) from the list. With the use of associated names, more accurate positions of the objects were obtained and the types of objects were found in Simbad. A few clearly non-blazar objects were left out, and optical magnitude selection was performed, as is described in \ref{sec:construction:GAIA}. An additional selection of objects to satisfy the isotropy criteria was performed on the basis of the flux of photons with energies between 1 and 100 GeV, $F_{\rm 1-100~GeV}$, as is described in Sec.~\ref{sec:criteria:latitude}, \ref{sec:criteria:sphere}. The criteria are satisfied for $F_{\rm 1-100~GeV}> 3.8\cdot10^{-10}$~cm$^{-2}$s$^{-1}$. The resulting histograms are presented in Fig.~\ref{Fig:Fermi_tests}. 
The criterion leaves 523 objects in the set, many of which are present in the VLBI sample as well. The combined sample consists of 651 sources.

\subsection{Optical magnitude selection}
\label{sec:construction:GAIA}
For both samples, we selected optically bright sources with the help of Gaia DR3 public data. We made use of the catalogs by \citet{OCARS,GAIA_1,GAIA_vizier,GAIA_2}. Correction for interstellar extinction was applied to the $G$-band magnitude with the help of the A2 model by \citet{GALExtin}, and the condition $G_{\rm corr}<18^{\rm m}$ was required for inclusion in our catalog. The value of $18^{\rm m}$ was fixed a priori, motivated by previous studies \citep{Gorbunov:2004}. 

\subsection{X-ray associations}
\label{sec:construction:Xray}
The construction of our main sample was discussed in Sec.~\ref{sec:construction:start}, \ref{sec:construction:GAIA}. However, we found it useful to include X-ray fluxes, if available, in the catalog. We used X-ray associations from several catalogs in the following order. Most of them were taken from the second ROSAT all-sky survey source catalogue \citep[2RXS][]{2RXS}. The absorption-corrected flux derived from the power law fit to the source spectrum, in erg/s/cm$^2$, was taken from that catalog as an X-ray flux of objects in the energy range $0.1-2.4$~keV. If not found in 2RXS, the source was searched for in the Fourth XMM-Newton Serendipitous Source Catalog, Thirteenth Data Release, or 4XMM-DR13 \citep{4XMM-DR13}. EPIC X-ray flux in the energy range $0.2 - 2.0$ keV was taken if no warning flags were set. The EPIC flux in each band is the mean of the band-specific detections in all of the cameras weighted by the errors. 
  
In case no data were obtained from 2RXS or 4XMM-DR13, we checked the second catalog of X-ray sources found in slew data taken by the European Space Agency's XMM-Newton Observatory, XMMSL2 or XMMSLEW, Version 2.0 \citep{XMMSL}, or the Chandra Source Catalog (CSC), the definitive catalog of X-ray sources detected by the Chandra X-ray Observatory \citep{2CXO}. Again, the X-ray flux in the energy range $0.2 - 2.0$ keV was obtained in case no warning flags were set.

Eventually, the flux in range $0.5 - 2.0$ keV  was obtained from the Swift AGN \& Cluster Survey (SACS) soft-band point source catalog \citep{SACS}.

\subsection{Object classification}
\label{sec:construction:class}
In the catalog, we also quote the classification of objects from SIMBAD \citep{Simbad}. SIMBAD utilizes International Virtual Observatory Alliance (IVOA) vocabulary of types of astronomical objects, ranging from stars to galaxies. According to its classification, all non-blazar objects are removed from the sample.

\subsection{Selection of confirmed BL Lacs}
\label{sec:construction:bll}
We took special care to collect a sample of confirmed BL Lacs, given their importance for cosmic ray studies discussed in the introduction. The definitions of BL Lacs vary in the literature. 
In particular, the SIMBAD classification of “BLLac” is not quantitatively defined. Therefore, we performed a manual selection of confirmed BL Lacs following the conventions of \cite{Veron} formulated by \citet{Veron-criteria}: the absence of emission lines with equivalent widths exceeding $5~\AA$. Making use of object-related references from NED, SIMBAD, and OCARS, we manually selected blazars satisfying these criteria. They include objects with unknown or photometric redshifts, objects with redshifts determined from the host galaxy features or from absorption lines, and objects for which numerical values of the emission-line widths, or the upper limits on them, were quoted in the literature. For this sample of confirmed BL Lacs, we tested the isotropy again with the criteria of Sec.~\ref{sec:criteria}. The resulting sample consists of 336 objects. 
The number of objects in the main blazar sample, classified as ``BLLac'' in SIMBAD, is 421.

\section{Description of the catalog}
\label{sec:catalog}
Both catalogs, of all blazars and BL Lacs, have been published in their entirety in a machine-readable format and are available with this paper. A portion of the main catalog is shown here; for guidance regarding its form and content, see Table~\ref{table:data_sample}. 

The first three columns include the J2000 name from RFC (J2000 name), the name from the 4FGL catalog (Fermi name), and the name from the corresponding X-ray catalog (X-ray name; not shown in the exemplary table but present in the machine-readable data). Coordinates in the equatorial system in decimal degrees (J2000) are also given. The column named “Radio flux” includes the mean flux density at 8~GHz from RFC, the “Gamma-ray flux” column includes the 4FGL flux in the energy range from 1 to 100 GeV, and the next column gives the X-ray flux in a corresponding energy range (from 0.2 to 2 keV or from 0.5 to 2 keV depending on the catalog; see Sec.~\ref{sec:construction:Xray} for details). The column “$G_{\rm corr}$” gives the corrected average magnitude in the G band. The terms used in column named “Type” represent type of object and were taken from the IVOA vocabulary, \url{http://www.ivoa.net/rdf/object-type/2020-10-06/object-type.html}.  The value of “-1” in any column of strings or “-1.0” in any column of numbers corresponds to a lack of data.
\begin{table*}
\caption{Exemplary isotropic sample data.}   
\begin{threeparttable}
\label{table:data_sample}  
\centering          
\begin{tabular}{ccccccccc}     
\hline\hline     
\cr  
J2000 name&Fermi name&RA,&DEC,&Radio flux&Gamma-ray flux,&X-ray flux& $G_{\rm corr}$, & Type
\cr &&deg&deg&Jy&cm$^{-2}$s$^{-1}$&erg\,cm$^{-2}$s$^{-1}$&mag \cr\\
\hline
\cr
  J2250+3824&4FGL\_J2250.0+3825&342.5240&38.4103&0.058&5.9526e-10&6.9032e-12&16.82&BLLac\\
  J2229-0832&-1&337.4170&-8.5485&2.455&-1.0&1.3030e-12&17.70&QSO\\
  J2030+1936& 4FGL\_J2030.9+1935&307.7381&19.6036&0.041&8.4981e-10&7.7035e-13&17.66&BLLac\\
  -1&4FGL\_J0331.3-6156&52.8269&-61.9247&-1.0&7.0249e-10&5.515e-12&17.80&BLLac\\
  J1728+5013&4FGL\_J1728.3+5013&262.0776&50.2196&0.094&1.6223e-09&3.969e-11&16.04&BLLac\\
  J1617-5848&-1&244.3245&-58.8022&1.572&-1.0&1.179e-13&17.18&Blazar\\
  J0521+2112&4FGL\_J0521.7+2112&80.4415&21.2143&0.354&1.162e-08&1.761e-10&15.81&BLLac\\
\cr
\hline   
\hline
\cr
\end{tabular}
\begin{tablenotes}
\item{N.B. The full table is available at the CDS.}
\end{tablenotes}
\end{threeparttable}
\end{table*}
The same information, except for the object type, is given for the BL Lac catalog in Table~\ref{table:bll_sample}.
\begin{table*}
\caption{Exemplary BL-Lac dataset.}    
\begin{threeparttable}
\label{table:bll_sample}   
\centering          
\begin{tabular}{ccccccccc}     
\hline\hline  
\cr
J2000 name&Fermi name&RA,&DEC,&Radio flux&Gamma-ray flux,&X-ray flux& $G_{\rm corr}$
\cr &&deg&deg&Jy&cm$^{-2}$s$^{-1}$&erg\,cm$^{-2}$s$^{-1}$&mag \cr\\
\hline
\cr
  J0001-0746&4FGL\_J0001.2-0747&0.3251&-7.77414&0.175&7.0245e-10&3.305e-11&16.38\\
  J0058-3234&4FGL\_J0058.0-3233&14.50929&-32.57243&0.131&1.3263e-09&-1.0&16.97\\
  J0209-5229&4FGL\_J0209.3-5228&32.34007&-52.48966&-1.0&2.2442e-09&1.249e-11&16.3\\
  J0217+0837&4FGL\_J0217.2+0837&34.32135&8.61775&0.65&1.8461e-09&4.94e-11&16.18\\
  -1&4FGL\_J0602.8-4019&90.71371&-40.31259&-1.0&1.2678e-09&6.389e-07&16.79\\
  J0617+5701&4FGL\_J0617.2+5701&94.32051&57.02123&0.308&1.4426e-09&-1.0&16.81\\
  J0147+4859&-1&26.9074&48.99375&0.133&-1.0&-1.0&17.14\\
\cr
\hline  
\hline
\cr
\end{tabular}
\begin{tablenotes}
\item{N.B. The full table is available at the CDS.}
\end{tablenotes}
\end{threeparttable}
\end{table*}

\section{Conclusions}
\label{sec:concl}
We present here a sample of blazars with extinction-corrected optical magnitudes, $G_{\rm corr}<18^{\rm m}$, satisfying the criteria of large-scale full-sky isotropy. These samples were selected from VLBI and gamma-ray catalogs and supplemented by X-ray associations. The catalog gives positions, radio, optical, X-ray and gamma-ray fluxes, and object types for 651 blazars uniformly distributed in the sky. A separate catalog gives 336 manually selected confirmed BL Lacs, satisfying the definition of \citet{Veron-criteria}.  

The catalog may be used in future searches for the sources of high-energy neutrinos and cosmic rays and in tests of several anomalies found in the astroparticle physics. In a separate paper \citep{iso-HiRes}, we apply this uniform sample to the HiRes cosmic rays and verify the conclusions of \citet{Gorbunov:2004,ST-anisotropy}. The
Telescope Array cosmic ray experiment is reaching the statistics required for the first tests of these HiRes results, and the present catalog may help in these coming tests. Other studies for which the isotropy of the sample is important may also benefit from the catalog.

\begin{acknowledgements}
We are indebted to Alexander Korochkin, Yury Kovalev, Mikhail Kuznetsov, Zinovy Malkin and Grigory Rubtsov for interesting and helpful discussions. 

This research has made use of the following public online astrophysical tools and services: the cross-match service, the VizieR catalogue access tool (doi:10.26093/cds/vizier, \cite{vizier_old}) and the SIMBAD database provided by the Strasbourg astronomical Data Center (CDS); of the NASA/IPAC Extragalactic Database (NED), which is operated by the Jet Propulsion Laboratory, California Institute of Technology, under contract with the National Aeronautics and Space Administration; of the data and software provided by the High Energy Astrophysics Science Archive Research Center (HEASARC), which is a service of the Astrophysics Science Division at NASA/GSFC; of data from the European Space Agency (ESA) mission {\it Gaia} (\url{https://www.cosmos.esa.int/gaia}), processed by the {\it Gaia} Data Processing and Analysis Consortium (DPAC, \url{https://www.cosmos.esa.int/web/gaia/dpac/consortium}, funded by national institutions, in particular the institutions participating in the {\it Gaia} Multilateral Agreement). 

This work was supported by the Russian Science Foundation, grant 22-12-00253.
\end{acknowledgements}
\bibliographystyle{aa} 
\bibliography{iso-catalog} 
\end{document}